\pdfoutput=1
\documentclass[conference,a4paper]{IEEEtran}
\makeatletter
\def\@IEEEpubidpullup{1.3\baselineskip}
\makeatother
\usepackage{jabbrv}
\usepackage[english]{babel}
\usepackage{amsmath,amssymb,xcolor,graphicx,fancyhdr}
\usepackage{tikz}
\usepackage{textcomp}
\usepackage{hyperref}
\usepackage{lipsum}
\newcommand\copyrighttext{%
  \footnotesize \textcopyright 978-1-4673-7545-0/15/\$31.00 2015 IEEE. Personal use of this material is permitted.
  Permission from IEEE must be obtained for all other uses, in any current or future
  media, including reprinting/republishing this material for advertising or promotional
  purposes, creating new collective works, for resale or redistribution to servers or
  lists, or reuse of any copyrighted component of this work in other works.}
\newcommand\copyrightnotice{%
\begin{tikzpicture}[remember picture,overlay]
\node[anchor=south,yshift=10pt] at (current page.south) {\fbox{\parbox{\dimexpr\textwidth-\fboxsep-\fboxrule\relax}{\copyrighttext}}};
\end{tikzpicture}%
}

{
\fancyhf{}
\fancyfoot[R]{}}

\newcommand{\assign}{:=}
\newcommand{\tmem}[1]{{\em #1\/}}
\newcommand{\tmop}[1]{\ensuremath{\operatorname{#1}}}
\newcommand{\tmrsup}[1]{\textsuperscript{#1}}
\newcommand{\tmstrong}[1]{\textbf{#1}}
\newcommand{\tmtextit}[1]{{\itshape{#1}}}

\begin{document}

\title{Go-Smart: Web-based Computational Modeling of Minimally Invasive Cancer
Treatments}

	\author{\IEEEauthorblockN{
	  Phil Weir\IEEEauthorrefmark{1},
	  Dominic Reuter\IEEEauthorrefmark{4},
	  Roland Ellerweg\IEEEauthorrefmark{4},
	  Tuomas Alhonnoro\IEEEauthorrefmark{3},
	  Mika Pollari\IEEEauthorrefmark{3},
	  Philip Voglreiter\IEEEauthorrefmark{2},\\
	  Panchatcharam Mariappan\IEEEauthorrefmark{1},
	  Ronan Flanagan\IEEEauthorrefmark{1},
	  Chang Sub Park\IEEEauthorrefmark{5},
	  Stephen Payne\IEEEauthorrefmark{5},\\
	  Elmar Staerk\IEEEauthorrefmark{4},
	  Peter Voigt\IEEEauthorrefmark{6},
	  Michael Moche\IEEEauthorrefmark{6},
	  and Marina Kolesnik\IEEEauthorrefmark{4}}
	  \IEEEauthorblockA{\IEEEauthorrefmark{1}
	    NUMA Engineering Services Ltd., Dundalk, Ireland}
	  \IEEEauthorblockA{\IEEEauthorrefmark{2}
	    Technical University of Graz, Graz, Austria}
	  \IEEEauthorblockA{\IEEEauthorrefmark{3}
	    Aalto University, Helsinki, Finland}
	  \IEEEauthorblockA{\IEEEauthorrefmark{4}
	    Fraunhofer Institute for Applied Information Technology, Germany}
	  \IEEEauthorblockA{\IEEEauthorrefmark{5}
	    University of Oxford, Oxford, United Kingdom}
	  \IEEEauthorblockA{\IEEEauthorrefmark{4}
	    Leipzig University, Germany}
	}
	\maketitle
	\copyrightnotice

\begin{abstract}
  The web-based Go-Smart environment is a scalable system that allows the
  prediction of minimally invasive cancer treatment. Interventional
  radiologists create a patient-specific 3D model by semi-automatic
  segmentation and registration of pre-interventional CT (Computed Tomography)
  and/or MRI (Magnetic Resonance Imaging) images in a 2D/3D browser
  environment. This model is used to compare patient-specific treatment plans
  and device performance via built-in simulation tools. Go-Smart includes
  evaluation techniques for comparing simulated treatment with real ablation
  lesions segmented from follow-up scans. The framework is highly extensible,
  allowing manufacturers and researchers to incorporate new ablation devices,
  mathematical models and physical parameters.
  
\end{abstract}

\section{Introduction}

Minimally invasive cancer treatments (MICTs) represent a growing body of
techniques for ablating cancerous tumors, avoiding major surgery. These MICT
modalities include, among others, radiofrequency ablation (RFA), where
percutaneous probes destroy tissue through Joule heating; microwave ablation
(MWA), where dielectric heating is used; cryoablation, where tissue is
lethally cooled; and irreversible electroporation (IRE), where the cell
membrane is destroyed by an electric field. In a rapidly growing discipline,
interventional radiologists (IRs) must remain informed about available
treatments, as their MICT experience heavily influences patient outcomes
{\cite{hildebrand-2006-influence}}. Moreover, the boundaries of tissue
necrosis are difficult to predict heuristically, and so experience is well
complemented by medical simulation. In addition to single-modality
offline workflows {\cite{rieder-2011-gpu,kerbl-et-al-2013-intervention}},
a multi-modality tool is thus required to flexibly compare a range of outcomes.

\begin{figure}
\centering

  \resizebox{200px}{!}{\includegraphics{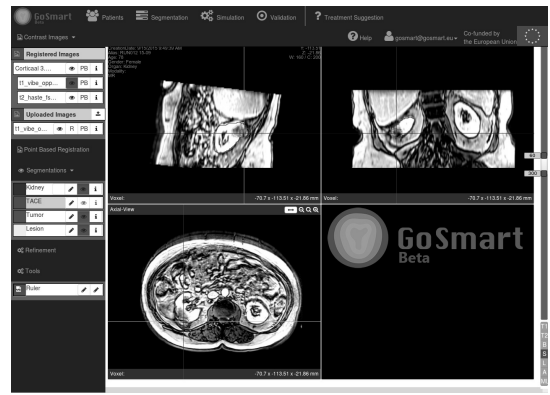}}
  \caption{\label{fig-screenshot}The 2D web-browser interface, showing image
  segmentation}
\end{figure}

\begin{figure}
\centering

  \resizebox{160px}{!}{\includegraphics{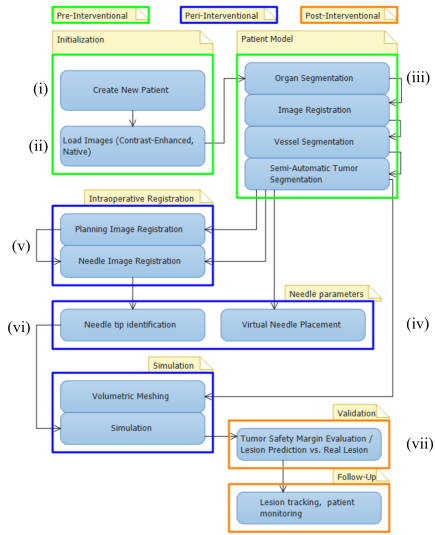}}
  \caption{\label{fig-computational-framework}Go-Smart framework: i)
  \tmtextit{Create project}; ii) \tmtextit{View data}; iii) \tmtextit{Segment}
  iv); \tmtextit{Place virtual needles}; v)\tmtextit{ }\tmtextit{Simulate};
  vi) \tmtextit{Register needles}; vii) \tmtextit{Validate}}
\end{figure}

The Go-Smart\footnote{Project Go-Smart is co-funded by the 7\tmrsup{th}
Framework Programme of the European Union under Grant Agreement No: 600641.}
project ({\tmem{http://gosmart-project.eu}}) seeks to fulfill this, expanding
on the RFA-specific IMPPACT\footnote{Project IMPPACT was co-funded by
the 7 \tmrsup{th} Framework Programme of the European Union under Grant
Agreement No: 223877.} project ({\tmem{http://imppact.eu}})
{\cite{kerbl-et-al-2013-intervention,payne-et-al-2011-image}}. A web-based
platform ({\tmem{http://smart-mict.eu}}) allows IRs to upload patient images,
and to plan, compare and validate treatment options (Fig.
\ref{fig-screenshot}). Establishing a complete environment has required
significant development, encompassing image segmentation, registration,
simulation, modeling and visualization, brought together within a
purpose-built scalable web architecture. Underpinning Go-Smart is the
principle that key MICTs share clinical and modeling commonalities. By
exploiting this, a generic computational framework (Fig.
\ref{fig-computational-framework}) has been defined and adapted to multiple
MICT modalities. A core feature of this framework is its extensibility;
mathematical models, numerical codes, equipment and even modalities may be
added through the interface, by independent researchers and manufacturers.

Pseudonymized validation data, for quantifying the performance of the image
manipulation and simulation components, is provided by clinical partners,
alongside a {\tmem{Case Report Form}} (CRF) containing all patient-specific
data needed for simulation. It is intended that the environment will support
independent evaluation of equipment, training of IRs, collaboration on
treatment planning and medical research.

\section{Usage}

\subsection{Clinicians}

Initially, IRs add a new patient with basic personal details (Fig.
\ref{fig-computational-framework}.i). Pre-interventional CT and/or MRI images
may be attached to the patient data and viewed in both 2D and 3D (Fig.
\ref{fig-computational-framework}.ii). Visible structures, such as the target
organ, tumors and vessels are segmented semi-automatically using a collection
of image analysis tools (Fig. \ref{fig-screenshot}; Fig.
\ref{fig-computational-framework}.iii), resulting in a patient-specific 3D
model of critical structures. Regions where trans-arterial chemoembolization
(TACE) has been applied, blocking blood supply to a tumor, may also be
segmented.

Before performing an intervention, IRs place virtual probes in the segmented
image, indicating the planned target (Fig.
\ref{fig-computational-framework}.iv). Equipment parameters and treatment
protocols are input, then a simulation is executed, which estimates the lesion
to be created (Fig. \ref{fig-computational-framework}.v). This step normally
takes 10-20 minutes, with variation for certain protocols, equipment and
modalities. After the procedure, IRs upload intra-operative scans, which are
registered to the pre-operative images, generating ablation probe coordinates
within the original patient model. The IR may re-run the simulation with these
real, measured probe locations (Fig. \ref{fig-computational-framework}.vi).
The physical lesion is segmented from follow-up scans performed 1-month
post-intervention, and quantitatively compared to the prediction (Fig.
\ref{fig-computational-framework}.vii).

\subsection{Researchers, developers and manufacturers}

For each treatment modality, manufacturers provide guidance for IRs, often as
complex algorithms, involving heating or cooling steps governed by time,
electrical impedance and/or reported temperature. The environment incorporates
these treatment protocols, simulating an IR or generator's behavior.

Researchers, developers and manufacturers may use the site's {\tmem{Developer
Corner}} to add such treatment protocols, numerical models (e.g. an Elmer
input file), or equipment. All models, probes, power generators, organs and
protocols have parameters, modifiable through this interface, which may also
be set to prompt for case-specific clinician input.

\section{Implementation}

\subsection{Distributed Architecture}

The Go-Smart distributed design is optimized for stability and scalability.
Allowing for computationally intensive workflow steps, processing is spread
between machines. The architecture is divided into independent,
separately-reusable components with clearly defined interfaces, as follows:
\begin{itemize}
  \item {\tmstrong{Clients}}: browser; {\tmem{Visapp}} (Go-Smart 3D Support);
  
  \item {\tmstrong{Servers}}: web server; image processing server;
  segmentation/registration server; simulation server.
\end{itemize}
Microsoft ASP.NET SignalR ({\tmem{http://signalr.net/}}) provides reliable,
rapid communication between services, allowing clients to interact directly
with each component. Together they build up a resilient architecture,
capable of handling substantial parallel workloads while retaining
satisfactory user experience.

The web application delivers the user interface, as well as managing
communication between all components. It supports multiple users and
scalability via a separately-hostable microservice. The database and CDM (Sec.
\ref{sec-cdm}) are also hosted here, as are patient images, 3D models and
simulation results.

\subsection{Browser Client}

A standard web browser is sufficient to access Go-Smart, providing users with
roaming access to patient data. The integrated viewer allows visual inspection
of all data generated during the workflow in a 2D, slice-based representation,
with axial, sagittal and coronal image windows. These can embed segmented and
simulated surfaces into an image (Fig. \ref{fig-screenshot}). Key tools, such
as image contrast windowing, provide a familiar radiology workspace,
comparable to established applications.

\subsection{Visapp Visualization Technology}

The optional client application provides seamlessly integrated 3D
visualization techniques, exploiting the local computing power of the client
PC to minimize response times. The client's embedded web browser replaces the
deactivated fourth view with a 3D render widget. Interaction with the
underlying web page is immediately reflected in the embedded 3D viewer when
required. The advanced 3D volume rendering techniques employed adjust to the
specifications of the local hardware. From basic direct volume ray casting
{\cite{levoy-1990-efficient}} to high-end global illumination techniques
{\cite{khlebnikov-2014-parallel}}, a variety of options are supplied to the
user. We incorporate both volumetric data and surface-based representation of
segmentation, simulation results and MICT probes, for exploring treatment
possibilities and evaluation purposes, supplementing the standard 2D views.

\subsection{Image Processing Server}

This component supports the client viewer, providing image re-slicing. All
image editing tools are integrated here, such as contrast setting,
segmentation editing and needle placement.

\subsection{Image Segmentation and Registration}

Within this component, the patient-specific 3D model is built using
purpose-built tools. First, separate images are registered into a common
coordinate system using multimodal semi-automatic registration. As well as
breath-hold compensation between consecutive contrast-enhanced sequences,
registration of intra-operative (needle) images and post-operative follow-up
validation images to pre-operative images is also implemented. The target
organ is automatically segmented through organ-specific tools, using a
rough decomposition of abdominal structures into semantic objects and
morphology-based segmentation. For certain organs and lesions, single-click
initialization is used for robustness, supplemented by `drawing' tools for
manual correction. Internal tubular structures, such as vessels, bile ducts and
bronchi, are extracted by an altering Hessian vessel model based
segmentation method.

\subsection{Clinical Domain Model (CDM)}\label{sec-cdm}

The Clinical Domain Model, a conceptual framework, gives separate identities
to components: numerical models, equipment, organs and protocols, each of which
maintains a set of parameters. Their relationships and allowed combinations are modified through a
\textit{Developer Corner} web-interface. Composition rules
enable interchangeability of individual components.

\subsection{Simulation Architecture (GSSA)}

The simulation architecture is a self-contained framework marshaled by a
CDM-produced XML file. It contains a fully-automated, configurable tool-chain,
with volumetric meshing by CGAL {\cite{cgal-2015-manual}} and simulation by
Elmer ({\tmem{https://www.csc.fi/web/elmer}}). New sandboxed numerical codes
may be added to GSSA using a Docker container ({\tmem{http://www.docker.com}})
and gluing Python module. An OpenFOAM container
({\tmem{http://www.openfoam.com}}) and a Python/FEniCS
{\cite{logg-2012-automated}} container have been implemented, allowing
interface users to run their own Python-based models.

\section{Mathematical Models}

Models used for simulating key modalities are outlined. For rigorous
discussion of the theory, we refer to {\cite{hall-2014-mathematical}}.

\subsection{Common Models of Thermal Modalities}

\subsubsection{Bioheat equation with perfusion term}

Following IMPPACT
{\cite{kerbl-et-al-2013-intervention,payne-et-al-2011-image}}, and Kr{\"o}ger
et al. {\cite{kroger-et-al-2006-numerical}}, a Pennes bioheat equation with
perfusion term is used for thermal modalities
{\cite{hall-2014-mathematical,hall-2015-cell}}. The governing volumetric
equation in the tissue is,
\begin{eqnarray}
  \rho c \partial_t T - k \nabla^2 T & = & Q_{\tmop{inst}} + Q_{\tmop{perf}},
\end{eqnarray}
where $\rho$, $c$, $k$ and $T$ \ are the density ($\mathrm{kg}\,\mathrm{m}^{-3}$),
specific heat capacity ($\mathrm{J}\,\mathrm{kg}^{-1}\,\mathrm{K}^{-1}$), thermal conductivity ($\mathrm{W}\,\mathrm{m}^{-1}\,\mathrm{K}^{-1}$) and
temperature of tissue ($\mathrm{K}$), respectively. $Q_{\tmop{inst}}$ is the
heat absorbed due to ablation ($\mathrm{W}\,\mathrm{m}^{-3}$). $Q_{\tmop{perf}}$ is proportional to deviation
from $310 \, \mathrm{K}$, with a tissue-type dependent factor.

\subsubsection{Cell death model}

For hyperthermia, a three-state cell death model is used
{\cite{hall-2015-cell,oneill-2011-three}}. Locally, cells are divided between
three states: Alive ($A$), Vulnerable ($V$) or Dead ($D$). At each point,
changes of state over time follow the rules,
\begin{equation}
	\begin{array}{llllll}
     A & \leftrightarrow & V & \rightarrow & D, & A + V + D = 1,
   \end{array} 
\end{equation}
according to coupled evolution equations with $T$-dependent coefficients. The
lesion is defined as $\Sigma \assign \{ D \geqslant 0.8 \}$.

\subsection{MICT-Specific Models}

\subsubsection{Microwave ablation}

This modality is modeled by coupling the above bioheat and death equations to
a reduced form of Maxwell's equations, using a transverse-magnetic (TM)
axisymmetric cylindrical solver with temperature-dependent electromagnetic
parameters
{\cite{hall-2014-mathematical,bertram-2006-antenna,ji-2011-expanded}}. The
field may be used to estimate the specific absorption rate of the tissue.

\subsubsection{Cryoablation}

A front-capturing multi-phase solver is applied to the Pennes equation to
ensure the accuracy of physical properties changing due to the expanding ice
ball. The {\tmem{effective heat capacity method}} is used, incorporating
latent heat of phase change into $c$ and adjusting $k$. This admits a
{\tmem{mushy}} transition region between solid and liquid states
{\cite{hall-2014-mathematical,deng-2004-modeling}}. The effective values are
inserted into the bioheat equation and the resulting nonlinear problem is
solved iteratively.

\subsubsection{Irreversible electroporation}

For each step, $i$, in the protocol, IRE is modeled using a simple electric
potential solver, with potential along the $i^{\tmop{th}}$ anode equal to the
$i^{\tmop{th}}$ potential difference, and zero potential along the
$i^{\tmop{th}}$ cathode. The final lesion is defined as an isovolume of the
local energy maximum over the whole protocol sequence
{\cite{hall-2014-mathematical,garcia-2014-numerical}}.

\subsubsection{Radiofrequency ablation}

Rather than simulating Joule heating for each execution of this modality, an
empirical approach is used, consisting of a summation of Gaussian functions
centered on suitably chosen points. This was validated during the IMPPACT
project, and avoids the oversized meshes required to capture $< 1 \,
\text{mm}$ diameter probe tines {\cite{payne-et-al-2011-image}}.

\section{Validation}

The clinical partners of the project supply validation data for the modalities
described in detail above, all uploaded and segmented by an experienced IR
through the standard interface. Validation measures include the
well-established average absolute error, $\alpha$ and target overlap, $\phi_S$
{\cite{klein-2009-evaluation}}. For a simulated lesion, $\Sigma$, and
segmented lesion, $S$, the absolute error at a point on $\partial S$ is the
minimum distance to a point on $\partial \Sigma$. $\alpha$ is then the surface
integral of these values divided by the area of $\partial S$. $\phi_S$ is
defined as the volumetric ratio, $| S \cap \Sigma | / | S |$.

The tool minimizes $\alpha$ over rigid motions of $S$, offsetting
post-operative registration errors. While this approach isolates inaccuracies
due to the normal clinical workflow, $\phi_S$, being independent, becomes a
more useful comparative measure. To demonstrate the process, sample cases are
presented in Tab. \ref{tab-validation-values}.

\begin{table}
\centering
\caption{\label{tab-validation-values}Validation measures for 8 cases (3
  s.f.)}

  \begin{tabular}{l}
    \begin{tabular}{ll|l|l|l|l}
      & {\tmstrong{Modality}} & Cryo.  & IRE & RFA & MWA\\
      \hline
      & {\tmstrong{Organ}} & Kidney & Liver & Liver & Liver\\
      1. & {\tmstrong{$\alpha$}} ($\text{mm}$) & 3.04 & 5.33 & 2.46 & 1.26\\
      & {\tmstrong{$\phi_S$}} & 0.953 & 0.357 & 0.701 & 0.593\\
      2. & {\tmstrong{$\alpha$}} ($\text{mm}$) & 1.95 & 3.28 & 2.01 & 1.41\\
      & {\tmstrong{$\phi_S$}} & 0.675 & 0.866 & 0.711 & 0.803
    \end{tabular}
  \end{tabular}
  
\end{table}

\begin{figure}
\centering

  \resizebox{!}{80px}{\includegraphics{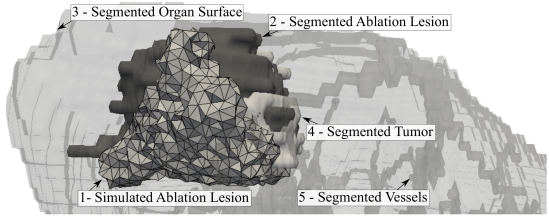}}
  \caption{\label{fig-rfa-1}Validation case RFA.1 showing simulated (1) and
  segmented (2) lesions, in a segmented liver (3) with tumor (4) and vessel
  structure (5)}
\end{figure}

While underestimation is shown by $\phi_S \ll 1.0$, overestimation, where $S
\ll \Sigma$, is indicated by large $\alpha$ and $\phi_S \lesssim 1.0$. From
Tab. \ref{tab-validation-values}, it is seen that the RFA samples are
particularly well-matched, with $\phi_S > 0.7$ and $\alpha < 2.5$ in both
cases (Fig. \ref{fig-rfa-1}). RFA benefits from longstanding project
experience and well-tested empirical parameters. IRE.1, in contrast, shows
$\alpha \gg 1$ and, for IRE.2, $\phi_S \ll 1$. Often measured IRE lesions are
small, and may require fuller modeling of device settings. Although MWA.2 is
adequate, MWA.1 underestimates, with $\phi_S < 0.6$. This may be related to
probe idiosyncrasies, fundamental to MWA, and rectified through further
tailored equipment modeling. Cryo.1 is overestimated, with $\alpha \gg 1$ and
$\phi_S \approx 1$, while Cryo.2 is underestimated, with $\phi_S \ll 1$. Many
treated kidney tumors are exophytic, with ablations close to the organ wall, a
limit of the simulation domain, $\Omega$. $| \Sigma |$ is thus sensitive to
needle placement error, arising from intra-operative to pre-operative image
registration. To resolve this, $\Omega$ may be extended beyond the organ.
Using Go-Smart's model extensibility, such changes may be made through the web
interface or back-end. As the overall validation error has two main
computational sources, image processing and simulation, we cannot reliably
distinguish their contributions. Yet some analysis is possible. Between liver
cases, the error shown is higher for MWA and IRE than for RFA, while image
processing steps are the same, implying simulation inaccuracy.

\section{Conclusion}

The Go-Smart web environment has been presented. It allows IRs to simulate
treatment for multiple MICT modalities and target organs, through a web-based
interface. The core simulation codes are then validated against IR-segmented
ablation lesions. As an open-ended system, researchers and manufacturers are
able to extend the framework with additional equipment, treatment protocols
and numerical models.

This framework represents a culmination of novel work
covering web infrastructure, image segmentation, image registration,
theoretical modeling, visualization techniques and simulation, both extending
and integrating underlying open source solvers. Potential
applications in e-health include international collaboration on
treatment planning, establishing a baseline for MICT training and independent
assessment of new equipment and techniques against an existing body of data.

\section*{Acknowledgments}

This research was funded by the European Commission, under Grant Agreement no.
600641, FP7 Project Go-Smart. The authors gratefully acknowledge the
significant contribution to this project made by our clinical partners at
Leipzig University, Medical University of Graz, Radboud University Medical
Centre Nijmegen and University Hospital Frankfurt.

\bibliographystyle{jabbrv_ieeetr}
\bibliography{all}
\end{document}